\documentclass[letterpaper]{jpconf}

\usepackage[dvips]{graphicx}
\usepackage{url,hyperref}

\makeatletter
  \newcommand\figcaption{\def\@captype{figure}\caption}
  \newcommand\tabcaption{\def\@captype{table}\caption}
\makeatother

\newcommand{\bvec}[1]{\ensuremath{\mbox{\boldmath $\mathbf{#1}$}}}

\begin{document}

\title{Looking for pentaquarks in Lattice QCD}

\author{George T.\ Fleming}

\address{Sloane Physics Laboratory, Yale University, 217 Prospect St,
         New Haven, CT 06520-8120 USA}

\ead{George.Fleming@Yale.edu}

\begin{abstract}
Pentaquark states in lattice QCD probably lie close in energy to two particle
scattering states. Correctly identifying the resonant state is a challenging,
yet tractable, problem given the terascale computing facilities available
today. We summarize the initial round of exploratory lattice calculations and
discuss what should be accomplished in the next round.
\end{abstract}

\section{\label{sec:introduction}Introduction}

The experimental evidence both for and against the existence of a narrow $S$=1
baryonic resonance around 1540 MeV is summarized elsewhere in these
proceedings \cite{Hicks:2004ge,Dzierba:2004db}.  If confirmed, this resonance
lies 5--10\% above the $KN$ scattering threshold with an allowed strong decay.
The implication for Lattice QCD theorists is that proper identification of
this resonance in Lattice QCD simulations \textsl{(even in the quenched
approximation)} should be substantially more difficult than the standard
calculation of the low-lying mass spectrum of hadrons stable against strong
decays.  Fortunately, techniques are being developed to study hadronic excited
states, scattering states and decays on the lattice.  That this narrow
resonance lies so close to the scattering threshold may make this one of the
easier scattering problems to study.

\begin{figure}[h]
\begin{minipage}{22pc}
  \begin{center}
    \tabcaption{\label{tab:summary}Current summary of lattice pentaquark
                spectroscopy (adapted from \cite{Sasaki:2004vz}).}
    \begin{tabular}{lllll}
      \br
      Ref. & $I; J^P$ & signal & $I$=0 parity & operator \\
      \mr
      \cite{Csikor:2003ng} & 0,1; $\frac{1}{2}^\pm$ & Yes & negative & other \\
      \cite{Sasaki:2003gi} & 0; $\frac{1}{2}^\pm$ & Yes & negative & diquark \\
      \cite{Mathur:2004jr} & 0,1; $\frac{1}{2}^\pm$ & No & N/A & $KN$ \\
      \cite{Chiu:2004gg} & 0; $\frac{1}{2}^\pm$ & Yes & positive & diquark \\
      \cite{Ishii:2004qe} & 0; $\frac{1}{2}^\pm$ & No & N/A & diquark \\
      \cite{Alexandrou:2004ws} & 0; $\frac{1}{2}^\pm$ & Yes & negative &
        diquark \\
      \cite{Negele:2004} & 0,1; $\frac{1}{2}^\pm$ & Yes & negative & many \\
      \cite{Takahashi:2004sc} & 0; $\frac{1}{2}^\pm$ & Yes & negative &
        $KN$, other \\
      \br
    \end{tabular}
  \end{center}
\end{minipage}\begin{minipage}{14pc}
  \begin{center}
    \includegraphics[width=0.9\textwidth]{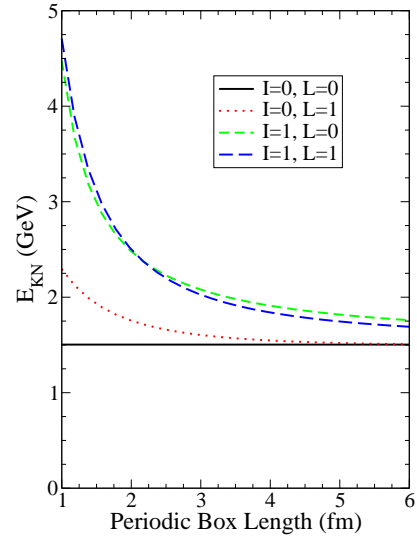}
    \caption{\label{fig:KNscattering}Volume dependence of $KN$ scattering.}
  \end{center}
\end{minipage}
\end{figure}

Initial Lattice QCD simulations studied the $S$=1 ground state energies for
various choices of total angular momentum $J$, isospin $I$ and parity $P$
\cite{Csikor:2003ng,Sasaki:2003gi,Mathur:2004jr,Chiu:2004gg,
Ishii:2004qe,Alexandrou:2004ws,Negele:2004,Takahashi:2004sc}.  From table
\ref{tab:summary}, we see that the lowest angular momentum ($J$=$\frac{1}{2}$)
isosinglet and isovector channels have been studied, but not isotensor
($I$=2).  In all these studies, the ground state energies were generally
consistent with a $KN$ threshold scattering state.  If the first excited state
energies were extracted and did not seem consistent with the next higher
scattering state, this indicated evidence for a possible pentaquark state.

The pentaquark operators used to create these quantum numbers on the lattice
generally had all five quarks at a single spatial point.  In this case, the
spatial symmetry enables a continuum-like operator construction with a
reasonable expectation of correctly identifying the $J^P$ of the operator.
Table \ref{tab:summary} indicates various strategies for constructing the
operators.

\section{\label{sec:scattering_states}Hadronic scattering states
        on the lattice}

Hadronic correlation functions in Lattice QCD behave as a sum of decaying
exponentials
\begin{equation}
\label{eq:correlator}
C_{ij}(t) = \left\langle
  \mathcal{O}_i(t)\ \mathcal{O}_j^\dagger(0)
\right\rangle = \sum_{n=0}^\infty \frac{Z_{in}}{\sqrt{2 E_n}}
\exp\left( - E_n t \right) \frac{Z_{jn}^*}{\sqrt{2 E_n}}
\end{equation}
where the operators $\mathcal{O}_i$ and $\mathcal{O}_j$ have the same quantum
numbers $(I, I_z, J, J_z, P, S, \cdots)$.  Since the statistical noise grows
exponentially in $t$ it is very difficult to reliably estimate the
contributions of even the first excited state ($n$=1) unless its energy $E_1$
is not much larger than the ground state $E_0$ and a judicious choice of
operators $\mathcal{O}_i$ can be found which have much better overlap with the
excited state ($\left| Z_{i1} \right| \gg \left| Z_{i0} \right|$).  This has
been the basic approach of most of the studies in table \ref{tab:summary}.

Assuming that an excited state energy can be reliably estimated, a resonance
can be distinguished from a scattering state by varying the spatial volume.
In a periodic box of size $L^3$, momentum components are quantized: $p_i = n_i
2 \pi / L$.  By varying $L$, energies of isolated moving hadronic resonances
vary predictably: $E_N^2(\vec{p}) = E_N^2(0) + (\vec{n} 2 \pi / L)^2$.
The energy of a two-particle scattering state will also vary according
to the scattering length $a_0$:
\begin{equation}
E_{KN}(\vec{p}, \vec{q}) = \left[ E_K(\vec{p}) + E_N( \vec{q}) \right]
\left[ 1 - \frac{2 \pi a_0}{E_K(\vec{p}) E_N(\vec{q}) L^3} \right] \left[
  1 - 2.834 \frac{a_0}{L} + 6.375 \left( \frac{a_0}{L} \right)^2
\right]
\end{equation}
Figure \ref{fig:KNscattering} illustrates the expected volume dependence of
$KN$ scattering for various isospin $I$ and orbital angular momentum $L$
channels where the scattering lengths have been experimentally determined.
However, the energies of isolated resonances at rest do not vary in
sufficiently large volumes where finite volume effects are not important.
Thus, scattering states and resonances can be distinguished in principle,
although a sufficiently wide range of volumes are seldom available in
practice.

\section{\label{sec:group_theory}Group-theoretic methods for hadron
         spectroscopy}

A better method for isolating hadronic excited states is a variational method
where a large basis of $N$ different operators $\mathcal{O}_i$ are identified
and an $N$$\times$$N$ Hermitian matrix of correlators $\bvec{C}(t)$ is
constructed.  The matrix elements of $\bvec{C}(t)$ are given by equation
(\ref{eq:correlator}).  If we make a \textsl{Schur decomposition} of
$\bvec{C}(t) = \bvec{Z} \bvec{\Lambda}^t \bvec{Z}^\dagger$ and truncate the
sum in equation (\ref{eq:correlator}) to the first $N$ terms, then we can make
an \textit{ansatz} that the $N$ eigenvalues of $\bvec{\Lambda}$ determine
the $N$ lowest energy levels: $\lambda_n = \exp(-E_n)$.  In practice,
it is better to find the $N$ solutions to the following equation,
called a \textsl{matrix pencil}:
\begin{equation}
\label{eq:pencil}
\bvec{C}(t_0)\ \bvec{z}_n = \lambda_n^{t-t_0} \bvec{C}(t)\ \bvec{z}_n,
\qquad \bvec{z}_m^\dagger\ \bvec{C}(t_0)\ \bvec{z}_n
= \lambda_m^{t_0} \delta_{mn}.
\end{equation}
The problem of isolating excited states is now reduced to finding a
sufficiently large basis of operators such that some linear combination will
overlap strongly with each state of interest.

Over the past few years, we have developed a group-theoretic technique for
constructing baryon operators that transform irreducibly under the lattice
symmetry group \cite{Edwards:2003mv,Basak:2003yd,Basak:2004hp,Basak:2004ib}.
We have extended our technique to pentaquarks by constructing operators
that are linear superpositions of gauge invariant terms of the form
\begin{equation}
\label{eq:elemental_operator}
\Phi_{\alpha i; \beta j; \gamma k; \mu m; \nu n}^{ABCDE} =
\mathcal{C}_{abcde}
\left( \widetilde{D}_i^{(p)} \widetilde{\psi} \right)_{a \alpha}^A
\left( \widetilde{D}_j^{(p)} \widetilde{\psi} \right)_{b \beta}^B
\left( \widetilde{D}_k^{(p)} \widetilde{\psi} \right)_{c \gamma}^C
\left( \widetilde{D}_m^{(p)} \widetilde{\psi} \right)_{d \mu}^D
\left( \widetilde{D}_n^{(p)} \widetilde{\overline{\psi}} \right)_{e \nu}^E
\end{equation}
where $A, B, C, D, E$ indices label quark flavor; $a, b, c, d, e$ indices
label color; $\alpha, \beta, \gamma, \mu, \nu$ are Dirac indices;
$\widetilde{\psi}$ ($\widetilde{\overline{\psi}}$) indicates a smeared quark
(antiquark) field; and $\widetilde{D}_j^{(p)}$ denotes the $p$-link covariant
displacement operator in the $j$-th direction.  The tensor
$\mathcal{C}_{abcde}$ indicates one of three linearly independent color
contractions needed to produce a color singlet.  For the displacements
considered here, the simplest choice $\mathcal{C}_{abcde} = \varepsilon_{abc}
\delta_{de}$ was sufficient to produce a complete set of operators.

For pentaquarks at rest, our operators transform as irreps of the
double-covered octahedral group $O_h$.  There are four two-dimensional
irreps $G_{1g}, G_{1u}, G_{2g}, G_{2u}$ and two four-dimensional irreps
$H_g$ and $H_u$.  Continuum spin assignments $J$ for lattice states
must be deduced from degeneracy patterns across different $O_h$ irreps
\cite{Johnson:1982yq}.

So far, we have constructed complete sets of two types of operators.  The
first type, called \textsl{single-site}, has all four quarks and the antiquark
at a single site.  The second type, called \textsl{singly-displaced}, has all
four quarks at one site and the antiquark displaced by $p$ sites along one of
the six spatial directions.  Table \ref{tab:linearly_independent} shows
the maximal number of linearly independent $I_z$=0, $S$=1 pentaquark
operators ($uudd\overline{s}$) for each isospin $I$ and operator type.

\begin{table}[h]
  \begin{minipage}{18pc}
    \caption{\label{tab:linearly_independent}Number of linearly independent
             $I_z$=0, $S$=1 pentaquark operators.}
    \begin{tabular}{llll}
      \br
      Operator Type    & $I$=0 & $I$=1 & $I$=2 \\
      \mr
      single-site      &   240 &   460 &   180 \\
      singly-displaced &  1440 &  2760 &  1080 \\
      \br
    \end{tabular}
  \end{minipage} \begin{minipage}{18pc}
    \caption{\label{tab:num_irreps}Number of irreducible representations
             for various $I_z$=0, $S$=1 pentaquark operators. Opposite
             parity irreps, \textit{e.g.}\ $G_{1g}$ and $G_{1u}$ occur
             with equal frequency.}
    \begin{tabular}{lllllll}
      \br
             & \multicolumn{3}{c}{single-site} &
                                       \multicolumn{3}{c}{singly-displaced} \\
      Irrep  & $I$=0 & $I$=1 & $I$=2 & $I$=0 & $I$=1 & $I$=2 \\
      \mr
      $G_1$  &    19 &    32 &    12 &    76 &   138 &    54 \\
      $G_2$  &     3 &     9 &     3 &    44 &    92 &    36 \\
      $H$    &    19 &    37 &    15 &   120 &   230 &    90 \\
      \br
    \end{tabular}
  \end{minipage}
\end{table}

Table \ref{tab:num_irreps} shows the maximal number of distinct irreducible
representations of $O_h$ that can be constructed for each operator type
and isospin $I$.  Irreps of opposite parity occur with equal frequency
so the number of $G_1$ irreps is the same for $G_{1g}$ and $G_{1u}$,
\textit{etc}.  Thus, if the observed $S$=1 resonance at 1540 MeV
is $I$=0 and $J$=$\frac{1}{2}$ then we can use up to 19+76=95 different
$G_1$ operators of either parity to construct our correlator matrices.
However, if neither the single-site nor the singly-displaced type
operators have a significant overlap with the resonance state, then
even more operators will need to be constructed using different
displacement patterns.

The next step in our program is to extend our existing software
which performs the needed Wick contractions to compute the elements
of baryon correlation matrices to include pentaquark operators.
We hope to report on initial quenched simulations in the near future.
We expect these pentaquark resonances will have a relatively large
spatial extent and heavy mass, so the use of improved actions
on anisotropic lattices will be important.  Once we have identified
linear combinations of operators which overlap strongly with the desired
resonances, it may be worthwhile to use the same operators to compute
correlation matrices on dynamical isotropic lattices as well.

\section{\label{sec:conclusion}Conclusion}

The lattice studies completed to date have most likely seen the threshold $KN$
scattering state in the $I$=0,1 $I_z$=0 $S$=1 channels.  In some cases, there
have been tantalizing hints of an excited state which have been interpreted as
either scattering states or resonances.  Until the lowest three or more
excited states can be cleanly extracted over a range of volumes, it remains
unlikely that lattice calculations can confirm or deny the existence of
pentaquark resonances in QCD.  The best method for extracting excited
states is the calculation of large correlator matrices.  The group-theoretic
operator construction technique can be used to produce the necessary
large basis of operators.

\ack{Portions of this work were performed in collaboration with members of the
hadron spectrum working group of the Lattice Hadron Physics Collaboration,
including S.~Basak, I.~Sato and S.~J.~Wallace (University of Maryland),
R.~G.~Edwards and D.~G.~Richards (Jefferson Lab), H.~R.~Fiebig (Florida
International University), U.~H.~Heller (American Physical Society) and
C.~Morningstar (Carnegie Mellon University).  This work was supported in part
by DOE contract DE-AC05-84ER40150 Modification No.~M175, under which the
Southeastern Universities Research Association (SURA) operates the Thomas
Jefferson National Accelerator Facility.}

\section*{References}

\bibliography{main}
\bibliographystyle{unsrt}

\end{document}